\documentclass{iau_FM} 
\usepackage{graphicx}
\usepackage{natbib}
\usepackage{aas_macros}
\usepackage{amsmath}
\usepackage[utf8]{inputenc}

\title[Improving Our Knowledge of the Solar Near-Surface Shear Layer] 
{Improving Our Knowledge of the Solar Near-Surface Shear Layer: The Special Case of the Leptocline}

\author[Jean-Pierre Rozelot, Alexander G. Kosovichev, Irina N. Kitiashvili]   
{Jean-Pierre Rozelot$^1$, Alexander Kosovichev$^{2,3}$, Irina Kitiashvili$^{3}$}
\affiliation{$^1$Universit\'e de la C\^ote d'Azur (emeritus), Grasse 06130, France\\
	email: {\tt jp.rozelot@orange.fr}\\
$^2$Center for Computational Heliophysics, New Jersey Institute of Technology, \\ University Heights, Newark, NJ 07102, U.S.A.\\
	email: {\tt alexander.g.kosovichev@njit.edu} \\
$^3$ NASA Ames Research Center, Moffett Field, CA 94035 U.S.A.\\
email: {\tt irina.n.kitiashvili@nasa.gov}}

\jname{Astronomy in Focus}
\editors{Worrall, D. Editor} 
\begin{document}

\maketitle

\noindent
V1: Figures and results unchanged. Minor changes: corrected typos; added references; acronym NSSL used instead full denomination.
\\

\begin{abstract}
The discovery of the solar activity cycle was linked from the outset to the observation of the temporal variability of sunspots, which we know to be the result of complex processes associated with the dynamics of inner layers. Numerous recent studies have highlighted changes in the Sun’s Near-Surface Shear Layer (NSSL),  pointing to the role of the leptocline, a shallow and sharp rotational shear layer in the top $\sim 8$ Mm. 
The leptocline, mainly characterized by a strong radial rotational gradient at middle latitudes and self-organized meridional flows, is the cradle of numerous phenomena: opacity, superadiabaticity, and turbulent pressure changes; the hydrogen and helium ionization processes; a sharp decrease in the sound speed; and, probably, variations of the seismic radius associated with a nonmonotonic expansion of subsurface layers with depth. In addition, the leptocline may play a key role in forming the magnetic butterfly diagram. Such results are a starting point for further systematic investigations of the structure and dynamics of this layer, which will lead to a better understanding of solar activity.
\end{abstract}

\firstsection 
\section{Introduction}

In recent years, numerous studies have focused on the physical conditions prevailing in the Sun's subsurface layers for at least two reasons. The first addresses the problem of how both the physical conditions in subsurface layers of the Sun and the nature of the magnetic flux tubes of active regions are reflected in the structure and behavior of these regions at the surface (e.g. Howard 1996; Choudhuri and Jha 2023; Rabello Soares et al.
2024; Kitchatinov 2023; Vasil et al. 2024). The second relates to differential rotation: the aim is to understand how the solar rotation, which is not uniform in latitude being faster at the equator than at the pole, varies in depth and time, a phenomenon known as “differential rotation” (for instance and references therein Javaraiah and Gokhale 2002; Javaraiah 2003; Tassoul 2000; Howe 2020).

Let us recall that the radiative interior of the Sun and its convective zone are separated, at a depth of around $0.7\,R_\odot$, by a thin layer ($\approx 0.05\,R_\odot$) at which the stratification changes rapidly from convective stability to marginal instability. This region shows a relatively sharp change between the solid-body rotation of the radiative interior and the differential rotation of the convection zone that Spiegel and Zahn (1992) termed the {\it tachocline}. 

Helioseismic studies, which are a powerful tool for probing the solar interior in three dimensions, show significant velocity variations in the near-surface layers (see for instance Komm 2022). Density decreases monotonically by several orders of magnitude; other physical conditions play an important role (see for Lefebvre et al. 2009). The treatment of the superadiabiatic region supposes a proper description of turbulent convection and detailed radiative energy transport and thermodynamic calculations, and we need to understand how the turbulent convection interacts with solar rotation. 

Furthermore, helioseismic studies have illustrated that the most significant changes with the solar cycle occur in a Near-Surface Shear Layer -hereinafter called NSSL,occupying around 5\% of the solar radius at the top of the convection zone. 
The velocity shear may convert a part of the poloidal magnetic field into the toroidal field, and, in addition to the global dynamo operating in the bulk of the convection zone (e.g. Pipin et al. 2023), the magneto-rotational instability may play a certain role (Vasil et al. 2024).

Helioseismic observations and numerical simulations reveal the existence of a shallow sub-surface  $\sim 8$\,Mm deep layer at the top of the NSSL (Kitiashvili et al. 2023; Rabello
Soares et al. 2024). By analogy to the tachocline, this layer is called “leptocline” (Godier
and Rozelot 2001), from the Greek “{\it leptos}”, thin and “{\it klino}”, tilt, or slope.  
This paper aims to present a new analysis of the gradient of solar rotation using global helioseismology data from the Solar and Heliospheric Observatory (SoHO) and Solar Dynamics Observatory (SDO) and highlight the role of the leptocline in our understanding of the structure and dynamics of the Sun and their variations with the solar activity cycles. 

\section{Radial Gradients of Solar Rotation: Tachocline and Leptocline}

The internal rotation of the Sun has been observed almost uninterruptedly since 1996 from two space missions, the Solar and Heliospheric Observatory (SoHO; Scherrer et al. 1995) and the Solar Dynamics Observatory (SDO; Scherrer et al. 2012) as well as from the ground-based Global Oscillation Network Group (GONG) network (Hill et al. 1996). The rotation rate is inferred by applying helioseismic inversion techniques to the rotational splitting of solar oscillation frequencies measured every 72 days from Doppler velocity images (Schou et al. 1997). 

\begin{figure}
	\begin{center}
		\includegraphics[width=\linewidth]{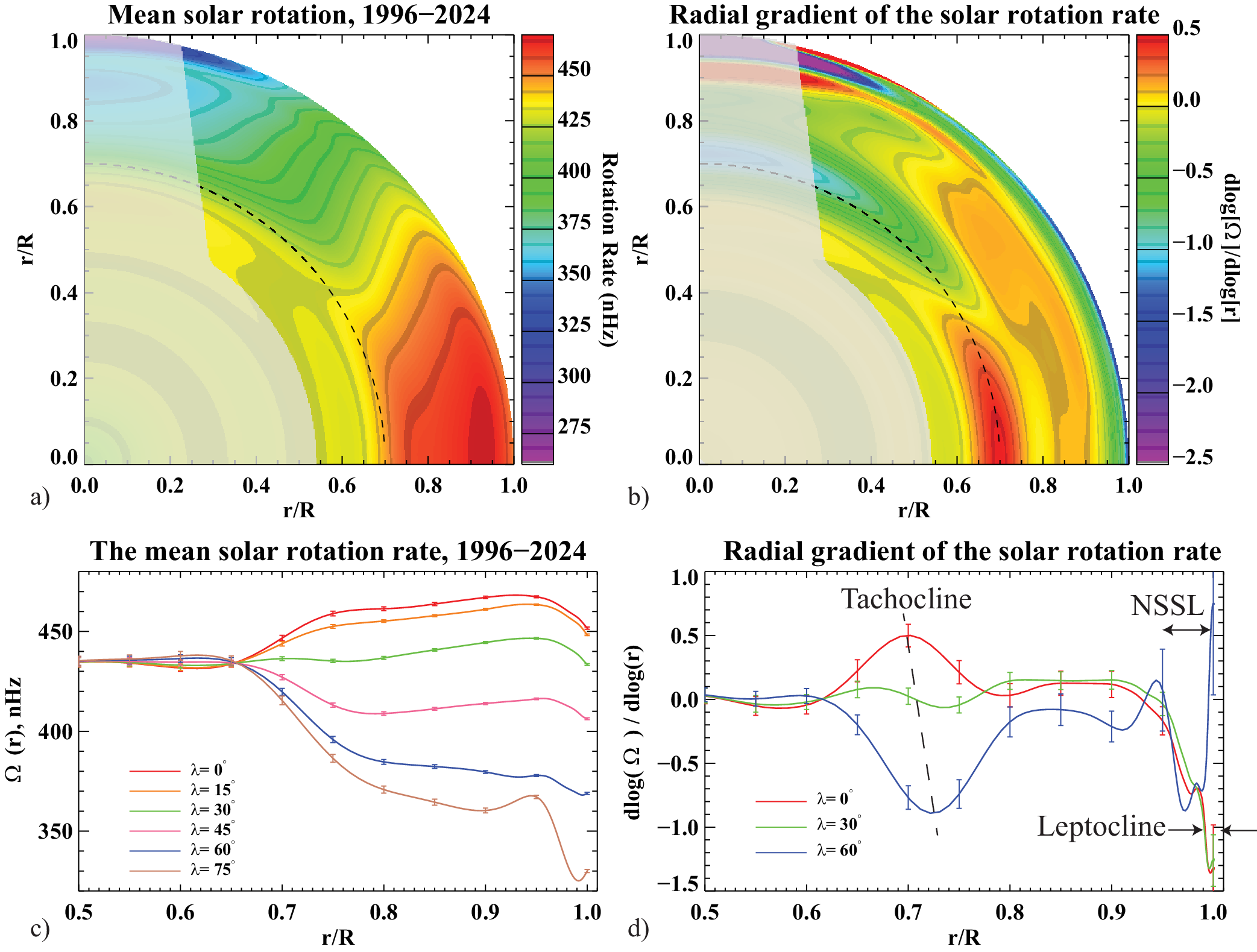} 
		\caption{
			The mean rotation rate, $\Omega$, and the radial gradient, $d\log\Omega/d\log r$, averaged over all SoHO/MDI and SDO/HMI measurements in 1996-2024, as a function of radius and latitude. a-b) cross-section views of the rotation rate and the gradient (the shaded region is where the inversion results are uncertain); c) the mean rotation rate at six latitudes indicated in the figure; d) the radial gradient as a function of radius at three latitudes indicated in the figure. 
		}\label{fig1}
	\end{center}
\end{figure}

Figure~\ref{fig1}a shows the solar rotation rate, $\Omega$, averaged over all available 141 inferences in 1996-2024 from SoHO and SDO, which are available from Joint Science Operation Center at Stanford\footnote{http://jsoc.stanford.edu} (for the methodology see Larson and Schou 2018). The gray areas in this image shows the regions of uncertainty where the averaging kernels of the helioseismic inversions are not well-localized (Schou et al. 1998). The logarithmic radial gradient of the rotation rate averaged over the same period is shown in Figure~\ref{fig1}b. The bottom panels (Fig.~\ref{fig1}c-d) show the radial dependencies of $\Omega$ and $d\log\Omega/d\log r$ for several latitudes. The error bars show the standard deviations of the weighted averages. 

These results are generally consistent with previous inferences of the internal solar rotation (e.g. Thompson et al. 1996; Kosovichev et al. 1997; Schou et al. 1998). In particular, the results reveal two zones of sharp gradients at the bottom of the convection zone (the tachocline) and its top (the NSSL). The gradient maximum values at the equator and 60 degrees latitude, connected by a dashed line in Fig.~\ref{fig1}d, indicated that the tachocline is deeper at the equator than at the high latitude by about 0.02-0.03\,$R_\odot$. This means the tachocline has a prolate shape as initially argued by Gough and Kosovichev (1995). 

The basic features of the internal solar rotation can be summarized as follows:
\begin{itemize}
\item Below 0.68\,$R_\odot$, the radiative interior rotates almost rigidly at a rate of about 430 nHz.
\item	The transition from the uniformly rotating radiation zone to a differentially rotating convection zone occurs in a thin layer from 0.68  to 0.73 $R_\odot$. This layer is called the tachocline.
\item	In the bulk of the convection zone (0.73 $R_\odot < r < 0.96 R_\odot$), the rotation rate varies strongly with latitude. The equator rotates about 30\% faster than the poles, from $\sim 460$ nHz at 0$^\circ$ latitude to $\sim 340$ nHz at 80$^\circ$ latitude.
\item	The contours of constant angular velocity are inclined by about 25$^\circ$ with respect to the rotational axis over a wide range of latitudes, i.e., rotation does not follow the Taylor-Proudman theorem.
\item	In a shallow layer, between 0.96 $R_\odot$ and 1 $R_\odot$, the rotation rate decreases by about 5\% at all latitudes, showing however a more complex behavior near the surface. This layer is called the NSSL.
\item	A substructure of the NSSL, the leptocline, located just below the surface, covers about 8 Mm in depth within the convection zone (0.985 $R_\odot < r < 1.0 R_\odot$).
\item	The leptocline unfolds an intricate behavior of the variation of the radial gradient,  $d\log\Omega/d\log r$, in latitude, depth, and in time.
\end{itemize}

\begin{figure}
	\begin{center}
		\includegraphics[width=\linewidth]{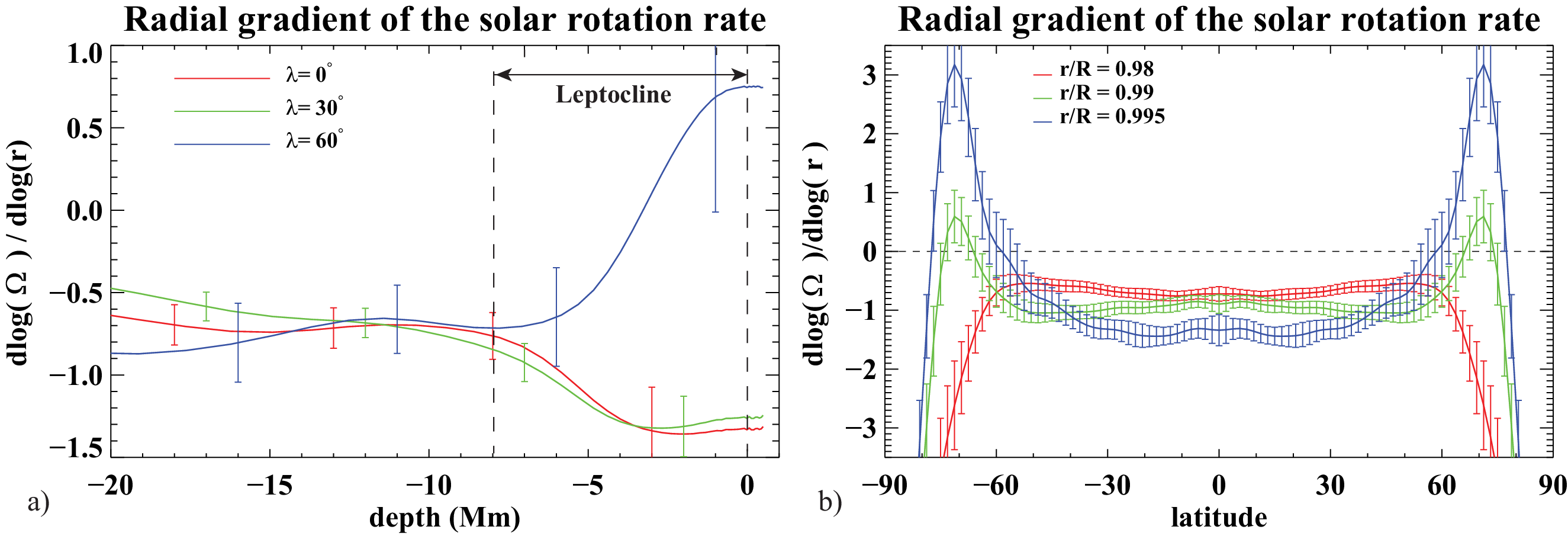}
	\caption{
		The radial gradient of the mean rotation rate, averaged over all SoHO/MDI and SDO/HMI measurements in 1996-2024, as a function of depth in the Near-Surface Shear Layer (NSSL) at three latitudes indicated in the figure.
	}\label{fig2}
\end{center}
\end{figure}

Figure~\ref{fig2} shows the radial gradient in the leptocline as a function of depth and latitude in more detail. The gradient that remains constant, $\simeq -1$, in the deep NSSL sharply increases its negative value to $\approx -1.5$ in the leptocline at the equator and low latitudes. But, at higher latitudes the gradient becomes positive, so that the rotation rate increases towards the surface. The gradient is constant in a top 2-3 Mm deep layer in these inversion results. However, we must emphasize that these are the global helioseismology inversions that include only the oscillation modes with an angular degree of up to 300. These data do not resolve sharp variations near the surface, smoothed with so-called averaging kernels (e.g. Schou et al. 1998). Therefore, the actual gradients of the rotation rate may be significantly larger than those revealed by the helioseismic inversions. 

\begin{figure}
	\begin{center}
		\includegraphics[width=0.65\linewidth]{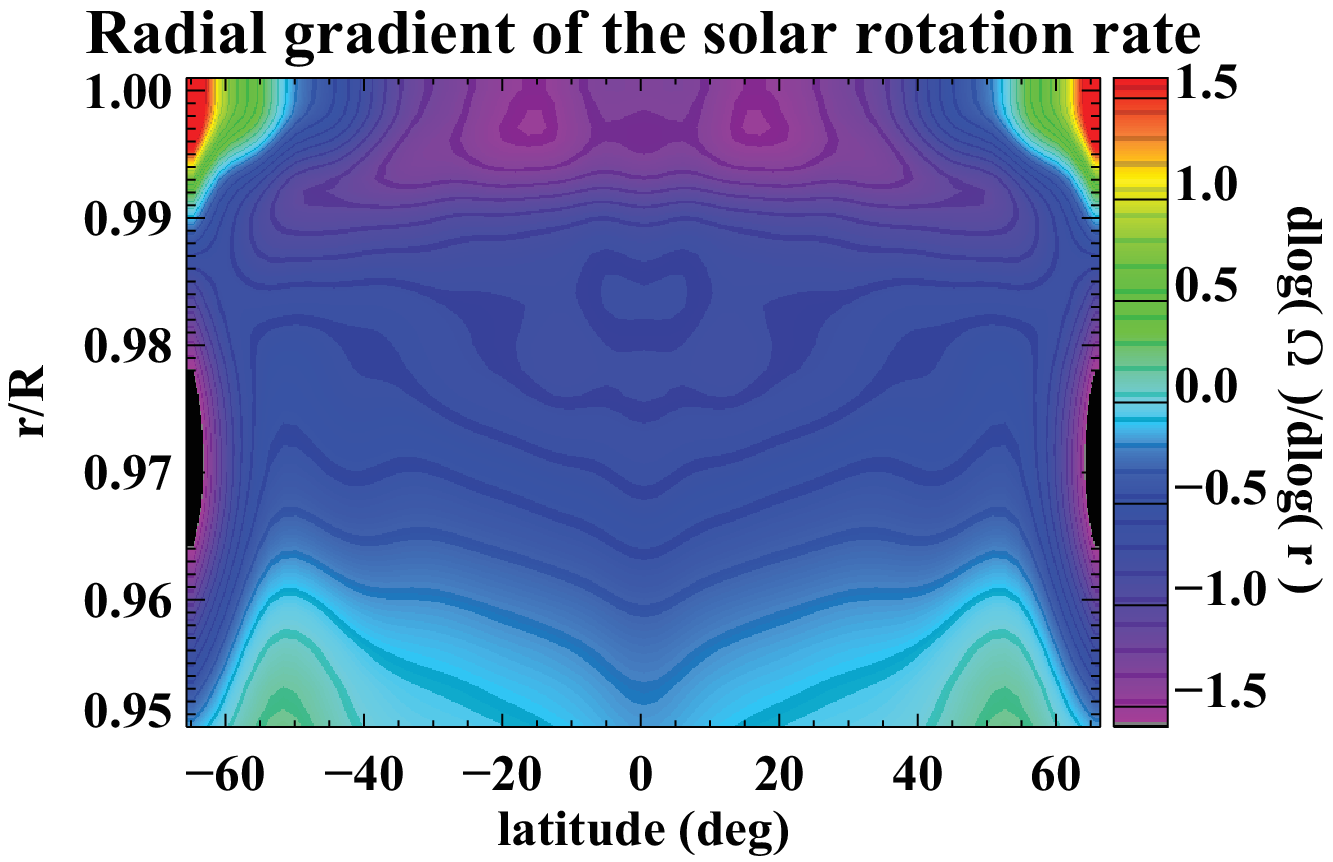}
	\caption{
	Variations of the rotational gradient, $d\log\Omega/d\log r$, in the leptocline a) with the depth below the solar surface at three latitudes, b) with latitude at three depths (shown in the figure); c) with latitude-radius diagram of the rotational gradient in the NSSL (0.96-1.0 $R_\odot$) and leptocline (0.99-1.0 $R_\odot$). 
	}\label{fig3}
\end{center}
\end{figure}

In fact, recent local helioseismology measurements based on the ring-diagram analysis that involved oscillation modes of high angular degree showed that the gradient at low latitudes can reach values of $\simeq -2.6$ at a depth of about 3~Mm and then reverse to smaller values at the surface (Komm 2022; Rabello Soares et al. 2024). The latitude-radius diagram of the mean rotational gradient shown in Figure~\ref{fig3} is qualitatively similar to the diagram obtained from the ring-diagram analysis (Figure 5 in Komm 2022). Nevertheless, there are significant differences, particularly in the latitudinal structure of the NSSL and the leptocline. For instance, the reversal of the gradient values from negative to positive at about 60$^\circ$ latitudes in the leptocline is prominent in the global helioseismology data. The ring-diagram analysis showed such reversal but in deeper layers below the leptocline. Such discrepancy must be resolved in future studies. 

\section{Solar-Cycle Variations of the Rotational Gradient}

Differential rotation also varies with time and typically reflects the solar cycle. After subtracting the mean rotation, the residual component revealed alternating zones of fast and slow flow bands, discovered by Howard and Labonte (1980) and called “torsional oscillations” because of their cyclic variations. The zonal flows originate at mid-latitudes and form two branches migrating toward the equator and polar regions just like the magnetic butterfly diagram but with the overlapping “extended” 22-year cycles (Wilson et al. 1988). 

  \begin{figure}[b]
	\begin{center}
		\includegraphics[width=0.65\linewidth]{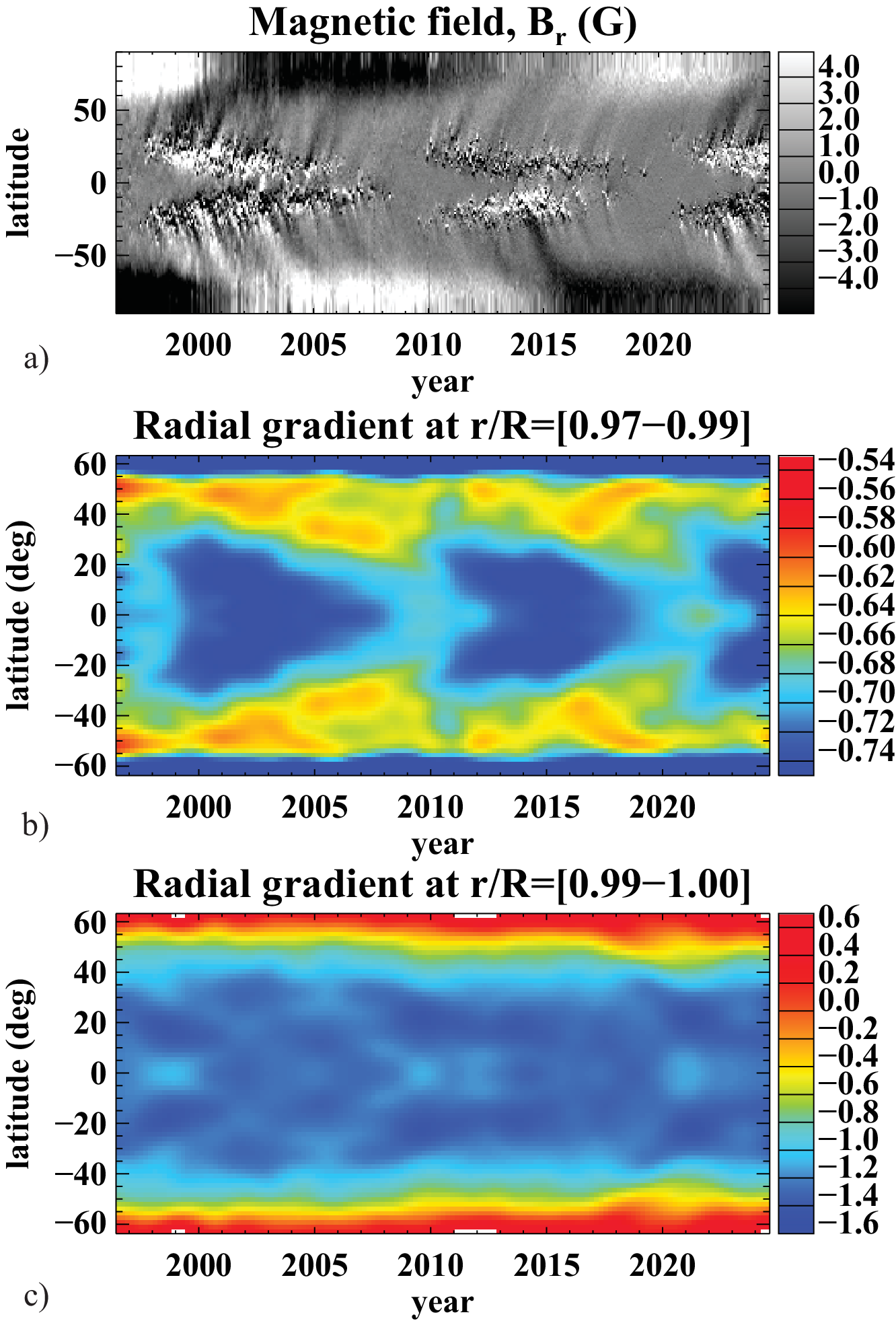}
	\caption{a) The magnetic butterfly diagram for the radial component of magnetic field calculated from the SoHO and SDO line-of-sight magnetic field data, assuming that the magnetic field
on the solar surface is predominately radial; b-c) the time-latitude diagrams for the rotational gradient, $d\log\Omega/d\log r$ below the leptocline ($r/R_\odot=0.97-0.99$) and in the leptocline ($r/R_\odot=0.99-1.00$) respectively. 
	}\label{fig4}
\end{center}
\end{figure}

Previous analyses of the internal rotation showed that the extended solar cycle represents the dynamo waves originating at the bottom of the convection zone and migrating towards the surface  (Kosovichev and Pipin 2019; Mandal et al. 2024). The dynamo model of Pipin and Kosovichev (2019) showed that these zone flows are due to the action of dynamo-generated magnetic fields and their effects on the convective heat transport and the meridional circulation in the solar convection zone. Both the observational data and the dynamo model show that the near-shear shear layer plays a key role in the formation of the magnetic butterfly diagram. The role of the leptocline in the solar dynamo has not yet been established. However, this shallow subsurface region is critical for the process of formation of sunspots and active regions. 

To illustrate the solar-cycle variations in the NSSL and leptocline, in Figure~\ref{fig4} we present the variations of the radial gradient 
as a function of time and latitude in two layers, just below the leptocline, at $r/R_\odot=0.97-0.99$ (or in the depth range of 7-21~Mm)
and in the leptocline (panel b), at $r/R_\odot=0.99-1.00$ (the corresponding depth range is 0-7 Mm; panel c). The comparison with the corresponding magnetic butterfly diagram (panel a) shows that below the leptocline, the negative gradient becomes stronger (dark blue areas in panel b) in the strong magnetic field regions migrating toward the equator and weaker in the high-latitude regions during the sunspot cycles. In the leptocline (panel c), the variations are less pronounced and have a more complicated structure, which, however, resembles the overlapping variation of the extended cycle of the torsional oscillations. In particular, the rotational gradient is stronger not only during the activity maxima but also during the activity minima when there are no strong magnetic fields on the solar surface. The gradient enhancement in the leptocline in quiet-Sun regions was previously noticed in the ring-diagram data (Komm 2022).    

\section{Variations of the Helioseismic Radius of the Sun With Respect to the Leptocline}

The solar-cycle variations of the Sun's rotation rate and its gradient in the NSSL and the leptocline are accompanied by structural changes related to the dynamo-generated magnetic fields emerging on the solar surface. The subsurface magnetic field has not been measured by helioseismology, although the first attempts to detect magnetic field signatures in the acoustic travel times have been made (Stefan and Kosovichev 2023). In general, the travel times and oscillation frequencies of acoustic waves (p-modes) depend on variations of both the magnetic field strength and temperature and their effects are not easy to separate in variations of the acoustic wave speed measured by helioseismology (Kosovichev et al. 2000; Dziembowski and Goode 2004).

It was noticed that the frequencies of surface gravity waves (defined as f-mode of solar oscillations) predominantly depend on the gravity acceleration on the solar surface and, thus, provide a measure of the solar radius, the so-called solar helioseismic (or seismic) radius of the Sun (Schou et al. 1997). Comparisons of the observed f-mode frequencies with the frequencies of the standard solar model  (Christensen-Dalsgaard et al. 1996) calibrated to the solar radius determined from optical observations showed a significant difference, indicating that the standard value of the solar radius must be reduced by about 300 km (Schou et al. 1997; Antia 1998). This result was later confirmed by analyses of p-mode frequencies (Kholikov and Hill 2008; Choudhuri and Jha 2023). A possible explanation is that the optical observations based on determining the position of the solar limb may be inaccurate due to the radiative transfer effects (Haberreiter et al. 2008) or incertitude due to differences in the definition of the solar radius (Rozelot et al. 2016).
  \begin{figure}
	\begin{center}
		\includegraphics[width=\linewidth]{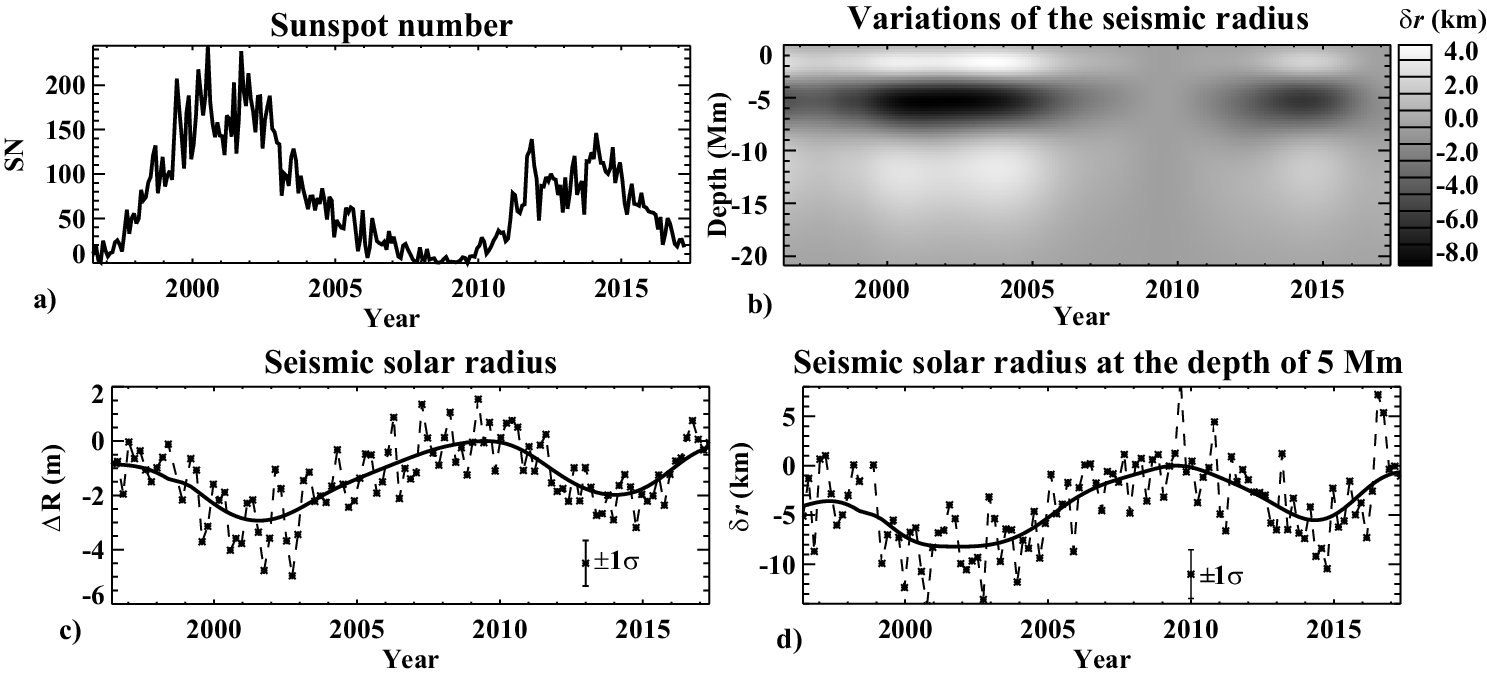}
	\caption{
		Variations of the seismic radius of subsurface layers during Solar Cycles 23 and 24: a) the sunspot number of these cycles; b) the time-depth diagram of subsurface displacements, $\delta r$, inferred from f-mode frequencies obtained from the SoHO and SDO data; c) the variations of the Sun's seismic radius obtained by averaging the displacements of the depth; d) the variations of the seismic radius ($\delta r$) in the leptocline, at a depth of 5~Mm. The data are adapted from Kosovichev and Rozelot (2018). Note that the averaged seismic radius changes in anti-correlation, meaning that it becomes smaller during the solar maxima, decreasing by about 3 km during the Cycle 23 maximum in 2000-2003, and by about 2 km during the Cycle 24 maximum in 2013-2015. 
	}\label{fig5}
\end{center}
\end{figure}

Further observations revealed variations of the seismic radius with the solar cycle, resulting in a reduction by several km with an increase in solar activity  (Dziembowski et al. 1998). Using nine years of data from SoHO, Lefebvre and Kosovichev (2005) established a variability of the helioseismic radius in antiphase with the solar activity, decreasing by about 2 km at the solar maximum (Figure~\ref{fig5}). Indeed, observations of the solar radiation integrated over
the entire solar spectrum (total irradiance), obtained by space-based experiments over the last decades, have demonstrated that total irradiance varies on time scales of minutes to the 11-year solar cycle (or more). As pointed out by Pap et al. (2001), if the central energy source remains constant while the rate of energy emission from the surface varies, there must be an intermediate reservoir where the energy can be stored or released depending on the variable rate of energy transport. The solar gravitational and magnetic fields could act as such a mechanism. If the energy is stored in this energy reservoir, it will result in a change in the solar radius, rending such faint variations plausible, by considering that the energy can be stored or released, depending on the variable rate of energy transport (Pap et al. 1998).

By applying a helioseismic inversion technique to the observed variations of f-mode frequencies, Kosovichev and Rozelot (2018) found that the seismic radius changes are associated with variations in the subsurface stratification (Figure~\ref{fig5}), with the strongest variations being just below the surface, around 0.995~R$_\odot$, that is about 3.5~Mm below the surface (Fig.~\ref{fig5}d). In addition, the radius of the deeper layers of the Sun, between 0.975 and 0.99\,$R_\odot$ changed in phase with the 11-year cycle. The variations of the displacement of the subsurface layers, $\delta r$, are illustrated in the time-depth diagram in Figure~\ref{fig5}b. 
Such variations in the leptocline stratification can be caused by subsurface magnetic fields and changes in the temperature distribution. 
 
 \section{Radiative Hydrodynamics Simulations of the Leptocline} 

\begin{figure}
	\begin{center}
		\includegraphics[width=\linewidth]{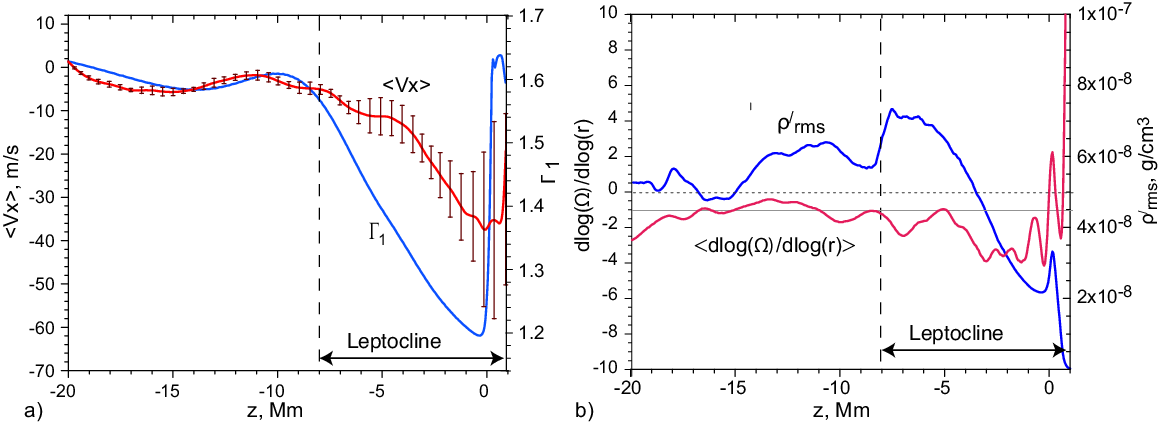}
		\caption{Mean radial profiles of a) deviations of the azimuthal flow speed from the imposed rotation rate at 30 degrees latitude (red curve) and the adiabatic index, $\Gamma_1$ (blue curve); b) the radial gradient of the rotation rate, defined as  ${\partial\log\Omega}/{\partial\log r}$ (red curve), and the RMS density perturbations, $\rho'_{rms}$ (blue curve). Radial profiles are obtained by averaging a 24-hour series of 3D simulation data horizontally over the simulation domain and in time. The vertical bars show $1\sigma$ flow velocity deviations from the mean values. The data in this figure are adapted from (Kitiashvili et al. 2023).
		}\label{fig6}
	\end{center}
\end{figure}

Kitiashvili et al. (2023) analyzed realistic 3D radiative hydrodynamics simulations of solar subsurface dynamics in the presence of rotation in a local domain 80 Mm wide and 25 Mm deep, located at 30 degrees latitude. The simulations revealed the development of a shallow 8-Mm deep substructure of the NSSL, characterized by strong turbulent flows and radial rotational gradient corresponding to the leptocline 
(Fig.~\ref{fig6}). 
It is located in the hydrogen ionization zone associated with enhanced anisotropic overshooting of convective flows (revealed by enhanced fluctuations of density, $\rho^\prime_{\rm RMS}$ in 
Fig.~\ref{fig6}b)
into a less convective unstable layer at a depth of about 8-12 Mm between the H\,I/He\,I and He\,II ionization zones, as illustrated by the adiabatic exponent $\Gamma_1$ in Fig.~\ref{fig6}a. 

The overshooting is characterized by intensified turbulent mixing. The azimuthal rotational velocity sharply decreases with depth by $\approx 38$~m/s in the leptocline. The gradient of rotation, $d\log\Omega/d\log r$, is about $-1$ in the NSSL below the leptocline and decreases to about -4 in the leptocline in agreement with observations. The simulations show a sharp increase of the gradient in a 2~Mm layer close to the surface in agreement with the helioseismic ring-diagram inferences (Rabello Soares et al. 2024).

\section{Conclusions}

In summary, the results of global and local helioseismology as well as 3D radiative hydrodynamic simulations show that the NSSL (Near Sub Surface Layer) occupying the top 15\% of the solar convection zone, the depth range $\approx 30-35$~Mm) has a distinct substructure, the leptocline, which is about 8~Mm deep and characterized by enhanced turbulent convection and a sharp increase in the rotational shear. 

The radial gradient of the solar rotation rate, $d\log\Omega/d\log r$, varies with the solar cycle. It is enhanced in regions of sunspot and active region formation. In middle and low latitudes, the gradient enhancements below the leptocline follow the magnetic butterfly diagram. However, in the leptocline, the latitudinal patterns of the enhanced gradient are more complicated, resembling the overlapping “extended'' solar cycles of the torsional oscillations. Curiously, the solar-cycle variations of the radial displacement of the subsurface layers, obtained from helioseismic inversion of f-mode (surface gravity waves) frequencies, are the strongest in the middle of the leptocline, at $\simeq 5$~Mm. The physical mechanism of the observed solar-cycle variations may be related to the accumulation of turbulent magnetic fields in this layer and associated changes in the temperature structure. 

High-resolution, realistic 3D hydrodynamic simulations reproduced the NSSL and the leptocline and showed that the rotational gradient might be stronger than found in the helioseismic inversions where the inferred rotation rate is smoothed within the averaging kernels. The simulations indicated that the origin of the leptocline is probably related to enhanced anisotropic turbulent convective downdrafts in the H\,I/He\,I ionization zone, which form an overshooting-type layer at the bottom of this zone at a depth of around 8~Mm (between the H\,I/He\,I and He\,II ionization zones), where turbulent mixing is intensified. It will be important to develop a synergy of helioseismic observations and numerical simulations for further understanding the complex turbulent physics of the leptocline and its role in the Sun's magnetic activity.

\medskip
\noindent
{\bf Acknowledgement.}
The work was partially supported by NASA grants: 
\\
\noindent
80NSSC19K0268, 80NSSC20K1320, and  80NSSC20K0602.

\bigskip

\noindent
{\bf References.}
{\small

Antia, H. M. 1998, Estimate of solar radius from f-mode frequencies. Astronomy and Astrophysics, 330, 336–340. doi:10.48550/arXiv.astro-ph/9707226.

Choudhuri, A. R. \& Jha, B. K. 2023,. The near-surface shear layer (NSLL) of the Sun: A theoretical
model. Proceedings of IAUS 365; pp. 4, [astro-ph.SR].

Christensen-Dalsgaard, J., Dappen, W., Ajukov, S. V., Anderson, E. R., Antia, H. M., Basu, S.,
Baturin, V. A., Berthomieu, G., Chaboyer, B., Chitre, S. M., Cox, A. N., Demarque, P.,
Donatowicz, J., Dziembowski, W. A., Gabriel, M., Gough, D. O., Guenther, D. B., Guzik,
J. A., Harvey, J. W., Hill, F., Houdek, G., Iglesias, C. A., Kosovichev, A. G., Leibacher,
J. W., Morel, P., Proffitt, C. R., Provost, J., Reiter, J., Rhodes, E. J., J., Rogers, F. J.,
Roxburgh, I. W., Thompson, M. J., \& Ulrich, R. K. 1996, The current state of solar
modeling. Science, 272(5266), 1286–1292.

Dziembowski, W. A. \& Goode, P. R. 2004, Helioseismic probing of solar variability: The formalism and simple assessments. ApJ, 600(1), 464–479.

Dziembowski, W. A., Goode, P. R., di Mauro, M. P., Kosovichev, A. G., \& Schou, J. 1998, Solar
cycle onset seen in Soho Michelson Doppler Imager seismic data. ApJ, 509(1), 456–460.

Godier, S. \& Rozelot, J. P. 2001, A new outlook on the ‘‘differential theory’’ of the solar
quadrupole moment and oblateness. Sol. Phys., 199, 217. doi:10.1023/A:1010354901960.

Gough, D. O. \& Kosovichev, A. G. An attempt to measure latitudinal variation of the depth of
the convection zone. In Hoeksema, J. T., Domingo, V., Fleck, B., \& Battrick, B., editors,
Helioseismology 1995,, volume 376 of ESA Special Publication, 47.

Haberreiter, M., Schmutz, W., \& Kosovichev, A. G. 2008, Solving the discrepancy between the
seismic and photospheric solar radius. ApJ, 675(1), L53.

Hill, F., Stark, P. B., Stebbins, R. T., Anderson, E. R., Antia, H. M., Brown, T. M., Duvall,
T. L., J., Haber, D. A., Harvey, J. W., Hathaway, D. H., Howe, R., Hubbard, R. P., Jones,
H. P., Kennedy, J. R., Korzennik, S. G., Kosovichev, A. G., Leibacher, J. W., Libbrecht,
K. G., Pintar, J. A., Rhodes, E. J., J., Schou, J., Thompson, M. J., Tomczyk, S., Toner,
C. G., Toussaint, R., \& Williams, W. E. 1996, The solar acoustic spectrum and eigenmode
parameters. Science, 272(5266), 1292–1295.

Howard, R. \& Labonte, B. J. 1980, The Sun is observed to be a torsional oscillator with a period
of 11 years. Astrophysical Journal Letters, 239, L33–L36. 10.1086/183286.

Howard, R. F. 1996, Solar active regions as diagnostics of subsurface conditions. Annual Review
of Astronomy and Astrophysics, 34, 75–109. doi:10.1146/annurev.astro.34.1.75.

Howe, R. Solar rotation. In Monteiro, M. J. P. F. G., Garc\'ia, R. A., Christensen-Dalsgaard,
J., \& McIntosh, S. W., editors, Dynamics of the Sun and Stars. Astrophysics and Space
Science Proceedings 2020,. Springer, Vol 57. doi:10.1007/978-3-030-55336-4 8.

Javaraiah, J. 2003, Long-term variations in the solar differential rotation. Solar Physics, 212(1),
23–49.

Javaraiah, J. \& Gokhale, M. K. 2002,. The Sun’s rotation. New York : Nova Science.

Kholikov, S. \& Hill, F. 2008, Sol. Phys., 251(1-2), 157–161. doi:10.1007/s11207-008-9205-9.

Kitchatinov, L. L. 2023, Origin of the near-surface shear layer of solar rotation. Astron. Lett.,
49(754-761), doi: 10. 1134/S106377372311004X.

Kitiashvili, I. N., Kosovichev, A. G., Wray, A. A., Sadykov, V. M., \& Guerrero, G. 2023,
Leptocline as a shallow substructure of near-surface shear layer in 3d radiative hydro-
dynamic simulations. Monthly Notices of the Royal Astronomical Society, 518(1), 504–512.\\
doi:10.1093/mnras/stac2946.

Komm, R. 2022, Radial gradient of the solar rotation rate in the near-surface shear layer of the
sun. Frontiers in Astronomy and Space Sciences, 9. doi:10.3389/fspas.2022.1017414. id.
428.

Kosovichev, A. G., Duvall, T. L. Jr., J., \& Scherrer, P. H. 2000, Time-distance inversion methods
and results - (invited review). Sol. Phys., 192, 159–176.

Kosovichev, A. G. \& Pipin, V. V. 2019, Dynamo wave patterns inside of the sun revealed by
torsional oscillations. ApJ, 871(2), L20.

Kosovichev, A. G. \& Rozelot, J. P. 2018, Cyclic changes of the Sun’s seismic radius. The
Astrophysical Journal, 861(2), id. 90. 5 pp. doi:10.3847/1538-4357/aac81d.

Kosovichev, A. G., Schou, J., Scherrer, P. H., et al. 1997, Structure and rotation of the solar interior: Initial results from the MDI medium-l program. Solar Physics, 170, 43.\\
doi:10.1023/A:1004949311268.

Larson, T. P. \& Schou, J. 2018, Global-mode analysis of full-disk data from the michelson doppler
imager and the helioseismic and magnetic imager. Sol Phys, 293(2), 29. doi:10.1007/s11207-
017-1201-5.

Lefebvre, S. \& Kosovichev, A. G. 2005, Changes in the subsurface stratification of the
sun with the 11-year activity cycle. The Astrophysical Journal, 633(2), L149–L152.
doi:10.1086/498305.

Mandal, K., Kosovichev, A. G., \& Pipin, V. V. 2024, Helioseismic properties of dynamo waves
in the variation of solar differential rotation. ApJ, 973(1), 36.

Pap, J., Kuhn, J., Fr\''ohlich, C. and Rozelot, J.P. 1998, in ''Proceedings of
 A crossroads for European Solar and Heliospheric Physics'', Tenerife, ESA, SP-417, 267.

Pap, J., Rozelot, J. P., Godier, S., \& Varadi, F. 2001, On the relation between total irradiance and radius variations. A\&A, 372, 1005–1018.

Pipin, V. V. \& Kosovichev, A. G. 2019, On the origin of solar torsional oscillations and extended
solar cycle. ApJ, 887(2), 215.

Pipin, V. V., Kosovichev, A. G., \& Tomin, V. E. 2023, Effects of emerging bipolar magnetic
regions in mean-field dynamo model of solar cycles 23 and 24. ApJ, 949(1), 7.

Rabello Soares, M. C., Basu, S., \& Bogart, R. S. 2024, Exploring the substructure of the near-
surface shear layer of the Sun. ApJ, 967(2), 143.

Rozelot, J.-P., Kosovichev, A., \& Kilcik, A. Solar radius variations: new look on the wavelength
dependence. In Kosovichev, A. G., Hawley, S. L., \& Heinzel, P., editors, Solar and Stellar
Flares and their Effects on Planets 2016,, volume 320 of IAU Symposium, pp. 342–350.

Scherrer, P. H., Bogart, R. S., Bush, R. I., et al. 1995, The solar oscillations investigation -
Michelson Doppler Imager. Solar Physics, 162(1-2), 129–188. doi:10.1007/BF00733429.

Scherrer, P. H., Schou, J., Bush, R. I., et al. 2012, Design and ground calibration of the Helio-Seismic and Magnetic Imager (HMI) instrument on the Solar Dynamics Observatory (SDO). Solar Physics, 275(1-2), 229–259. doi:10.1007/s11207-011-9842-2.

Schou, J., Antia, H. M., Basu, S., Bogart, R. S., Bush, R. I., Chitre, S. M., Christensen-
Dalsgaard, J., Di Mauro, M. P., Dziembowski, W. A., Eff-Darwich, A., Gough, D. O.,
Haber, D. A., Hoeksema, J. T., Howe, R., Korzennik, S. G., Kosovichev, A. G., Larsen,
R. M., Pijpers, F. P., Scherrer, P. H., Sekii, T., Tarbell, T. D., Title, A. M., Thompson,
M. J., \& Toomre, J. 1998, Helioseismic studies of differential rotation in the solar envelope
by the solar oscillations investigation using the Michelson Doppler Imager. ApJ, 505(1),
390–417.

Schou, J., Kosovichev, A. G., Goode, P. R., \& Dziembowski, W. A. 1997, Determination of the
sun’s seismic radius from the SoHO Michelson Doppler Imager. Astrophysical Journal Letters,
489, L197. doi:10.1086/31678.

Spiegel, E. A. \& Zahn, J. P. 1992, The solar tachocline. Astronomy and Astrophysics, 265,
106–114.

Stefan, J. T. \& Kosovichev, A. G. 2023, Exploring the connection between helioseismic travel
time anomalies and the emergence of large active regions during solar cycle 24. ApJ, 948(1), 1.

Tassoul, J. L. 2000, Stellar rotation. Cambridge University Press, 240 p.

Thompson, M. J., Toomre, J., Anderson, E. R., Antia, H. M., Berthomieu, G., Burtonclay,
D., Chitre, S. M., Christensen-Dalsgaard, J., Corbard, T., De Rosa, M., Genovese, C. R.,
Gough, D. O., Haber, D. A., Harvey, J. W., Hill, F., Howe, R., Korzennik, S. G., Kosovichev,
A. G., Leibacher, J. W., Pijpers, F. P., Provost, J., Rhodes, E. J., J., Schou, J., Sekii, T.,
Stark, P. B., \& Wilson, P. R. 1996, Differential rotation and dynamics of the solar interior.
Science, 272(5266), 1300–1305.

Vasil, G. M., Lecoanet, D., \& Augustson, K. 2024, The solar dynamo begins near the surface.
Nature, 629(8013), 769–772. doi:10.1038/s41586-024-07315-1.

Wilson, P. R., Altrocki, R. C., Harvey, K. L., Martin, S. F., \& Snodgrass, H. B. 1988, The
extended solar activity cycle. Nature, 333(6175), 748–750.
}
\end{document}